\documentclass[review=false]{actapoly}

\usepackage[normalem]{ulem}
\usepackage{bm}

\newcommand{\Fref}[1]{Figure~\ref{#1}}
\newcommand{\Eref}[1]{Eq.~\eqref{#1}}

\newcommand{\de}[1]{\,{\mathrm d}#1} 
\newcommand{\trn}{{\sf ^T}}  

\newcommand{\q}{\bm{q}} 
\newcommand{\lam}{\bm{\lambda}} 
\newcommand{\half}{{\textstyle\frac{1}{2}}}
\newcommand{\quar}{{\textstyle\frac{1}{4}}}
\newcommand{\discr}{_{\mathrm d}}

\newcommand{\secref}[1]{Section~\ref{#1}}
\newcommand{\figref}[1]{Figure~\ref{#1}}
\newcommand{\tabref}[1]{Table~\ref{#1}}
\newcommand{\trial}[1]{\widehat{#1}}
\newcommand{\uV}{\bm{u}}
\newcommand{\eps}{\bm{\varepsilon}}
\newcommand{\sig}{\bm{\sigma}}
\newcommand{\nablas}{\bm{\nabla}_\mathrm{s}}

\begin{document}

\title[Newmark algorithm for dynamic analysis of viscoelastic materials]
{Newmark algorithm for dynamic analysis with Maxwell chain model}

\correspondingauthor[J. Schmidt]{Jaroslav Schmidt}{ctu}{Jaroslav.Schmidt@cvut.cz}
\author[T. Janda]{Tom\'a\v s Janda}{ctu}
\author[A. Zemanov\'a]{Alena Zemanov\'a}{ctu}
\author[J. Zeman]{Jan Zeman}{ctu}
\author[M. \v Sejnoha]{Michal \v Sejnoha}{ctu}

\institution{ctu}{Department of Mechanics, Faculty of Civil Engineering, Czech Technical University in Prague, Th\'akurova 7, Praha 6, Czech Republic}

\begin{abstract}
This paper investigates a time-stepping procedure of the Newmark type for dynamic analyses of viscoelastic structures characterized by a generalized Maxwell model.
We depart from a scheme developed for a three-parameter model by Hatada et al.~\cite{hatada2000dynamic}, which we extend to a generic Maxwell chain and demonstrate that the resulting algorithm can be derived from a suitably discretized Hamilton variational principle. 
This variational structure manifests itself in an excellent stability and a low artificial damping of the integrator, as we confirm with a mass-spring-dashpot example.
After a straightforward generalization to distributed systems, the integrator may find use in, e.g., fracture simulations of laminated glass units, once combined with variationally-based fracture models.
\end{abstract}


\maketitle

\section{Introduction}

The motivation of this work comes from the field of dynamics of laminated glass structures. These sandwich structures consist of multiple glass layers connected with transparent polymer interlayers. Combining stiff, brittle glass with compliant viscoelastic polymers enhances structural safety, but the through-thickness heterogeneity renders mechanics of laminated glass structures intricate, e.g.~\cite{Haldimann:2008:SUG}. In particular, time- and temperature-dependent interlayer properties must be accounted for even in quasi-static analyses, e.g.,~\cite{duser_alex_van_analysis_1999,galuppi_design_2013,zemanova_comparison_2017,zemanova_layer-wise_2018} and references therein.

Earlier studies~\cite{Andreozzi:2014:DTT,Shitanoki:2014:PNM,Mohagheghian:2017:QSB,Hana2019} have shown that the response of commonly used interlayer materials can be captured well with the Maxwell chain model combined with the time-temperature superposition principle. Because the viscoelastic model concurrently predicts material damping, vibrations of laminated glass structures can be described more accurately than in conventional structural analyses that mostly employ the Rayleigh damping, e.g.~\cite[Section 12.5]{clough2003dynamics}. This added value has been addressed in detail for free vibrations of laminated glass units, e.g.~\cite{koutsawa2007static,aenlle_frequency_2013,zemanova_modal_2018}; an extension towards the response under general dynamic loads requires the development of dedicated time-stepping schemes that are in the focus of the current work.

\paragraph{Related work.} Dynamics of viscoelastic solids described by the Maxwell chain model leads to the system of initial value problems coupling the equation of motion with the local evolution of constitutive variables, see \secref{sec:governing} for illustration. Because numerically integrating the full system would be costly, we will follow an alternative approach in which only the equations of motion are solved approximately, whereas the evolutionary constitutive equations are resolved in the closed form, leading to an inexpensive update formulas for internal variables entering the equations of motion. This approach has been pioneered for quasi-static problems by Zienkiewicz~et~al.~\cite{zienkiewicz_numerical_1968}; see also~\cite[Section~5.2]{bavzant2018creep} for a comprehensive review. To the best of our knowledge, Hatada et al.~\cite{hatada2000dynamic} were the only ones who used this strategy in dynamics, although no reference to the original work~\cite{zienkiewicz_numerical_1968} was made. In particular, they developed a Newmark-type~\cite{Newmark1959} algorithm for the three-parameter Maxwell model and used it to predict the response of planar frames to earthquake loading.

\paragraph{Novelty.} Our work further develops the contribution~\cite{hatada2000dynamic} in three aspects. First, in \secref{sec:discretization}, we present a compact derivation of the Newmark scheme for a generic Maxwell chain, closely following the original exposition~\cite{zienkiewicz_numerical_1968}. Second, in \secref{sec:vi}, we show that the algorithm can be interpreted as a variational integrator~\cite{Kane2000}, in the sense that it can be derived from the Hamilton variational principle combined with suitable time discretization. The variational structure endows the integrator with good numerical stability and low numerical dissipation, as demonstrated in \secref{sec:example_discrete_problem} with selected examples. Moreover, the scheme can be easily combined with variational approaches to fracture, e.g.,~\cite{bourdin2008variational,buliga_hamiltonian_2009,bourdin_time-discrete_2011}, which is of independent interest when simulating the behavior of laminated glass under impact loads. Third, in \secref{sec:example_continuum}, we outline how to extend the algorithm to a continuum formulation and complement the theoretical considerations with an illustrative 3D finite element simulation. 

\paragraph{Notation.} We employ the conventional notation through the text, in which scalar quantities are denoted by a plain font, whereas bold-face letters indicate vectors or higher-order tensors. Additional nomenclature is introduced when needed.

\section{Newmark method}\label{sec:Newmark}

\begin{figure}[h]
\begin{center}
\def\svgwidth{80mm}
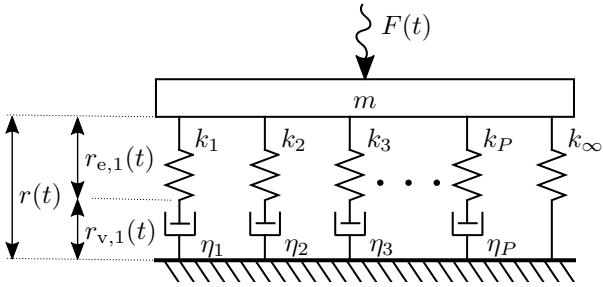
\caption{Scheme of the single-degree-of-freedom viscoelastic dynamic problem.}
\label{fig:maxwellP}
\end{center}
\end{figure}

In this section, we analyze a single degree of freedom (SDOF) model of a mass supported with a Maxwell chain, consisting of the parallel connection of an elastic spring and multiple spring-dashpot cells, see~\Fref{fig:maxwellP} for illustration and, e.g.,~\cite[Section~A]{bavzant2018creep} for further details. In particular, in Section~\ref{sec:governing} we review the equations of motion, which we subsequently discretize with the average acceleration version of the Newmark method~\cite{Newmark1959} in Section~\ref{sec:discretization}.

\subsection{Governing equations}\label{sec:governing}

As follows from the scheme in~\Fref{fig:maxwellP}, the problem under consideration is specified with the time-dependent load $F(t)$, the particle mass $m$ and the Maxwell chain model parameters: stiffness of the elastic spring $k_\infty$, spring stiffness $k_p$ and damper viscosity $\eta_p$ of the $p$-th Maxwell cell; $P$ stands for the number of Maxwell cells.

Equilibrium of the forces acting on the mass requires
\begin{eqnarray}
m \ddot{r}(t) + k_\infty r(t) + \sum_{p=1}^P f_p(t) = F(t),
\label{eq:equilibrium}
\end{eqnarray}
where $r$ denotes the displacement of the mass, $\ddot{r}$ its acceleration, and $f_p$ the restoring force of the $p$-th cell.

For the $p$-th Maxwell cell, the displacement $r$ splits into an elastic part of the spring $r_{\mathrm{e},p}$ and a viscous part of the damper $r_{\mathrm{v},p}$:
\begin{eqnarray}
r(t) = r_{\mathrm{e},p}(t) + r_{\mathrm{v},p}(t);
\label{eq:evdisp}
\end{eqnarray}
recall \Fref{fig:maxwellP}. The restoring force of the $p$-th Maxwell cell satisfies
\begin{eqnarray}
f_p(t) = k_p r_{\mathrm{e},p}(t) = \eta_p \dot{r}_{\mathrm{v},p}(t),
\label{eq:force_p_equil}
\end{eqnarray}
because of the serial arrangement of the spring and damper in the cell. Differentiating~\eqref{eq:evdisp} with respect to time and using~\eqref{eq:force_p_equil}, we obtain
\begin{align}\label{eq:rdvisco}
\frac{\dot{f}_p(t)}{k_p} + \frac{f_p(t)}{\eta_p} 
=
\dot{r}(t). 
\end{align}

In summary, the motion of SDOF model is described with the coupled system $(P+1)$ ordinary differential equations (ODEs)~\eqref{eq:equilibrium} and~\eqref{eq:rdvisco}, complemented with the initial conditions
\begin{align}\label{eq:initial_conditions}
r(0) = r_0, 
\quad
\dot{r}(0) = v_0,
\quad
f_p(0) = f_{p, 0},
\end{align}
where $r_0$ and $v_0$ stand for the initial mass displacement and velocity, $f_{p, 0}$ is the initial force in the $p$-th Maxwell cell, and $p = 1, \ldots, P$.

\subsection{Discretization}\label{sec:discretization}
The time interval of interest $\langle0,
T\rangle$ is divided into $(N+1)$ time instants $0 = t_0 < t_1 <
t_2 < \cdots < t_{N-1} < t_N = t_{\mathrm{max}}$; for notational simplicity, we assume equidistant partitioning of the constant time step $\Delta t = t_{i+1} - t_i$.

Considering the Newmark integration scheme~\cite{Newmark1959} with an average acceleration $(\ddot{r}_i+\ddot{r}_{i+1})/2$ on the time interval $\langle t_i, t_{i+1}\rangle$, the velocity and displacement within the interval varies as
\begin{subequations}
\label{eq:Newmark_kinematics}
\begin{align}
\dot{r}(t_i + \tau) &= \dot{r}_i
+ \half
\left(
\ddot{r}_i
+ \ddot{r}_{i+1}
\right)
\tau,\label{eq:vapprox}\\
r(t_i + \tau) &= r_i
+ \dot{r}_i
\tau + {\textstyle \frac{1}{4}}
\left(
\ddot{r}
+ \ddot{r}_{i+1}
\right)\tau^2,\label{eq:rapprox}
\end{align}
\end{subequations}
where $\tau \in \langle 0, \Delta t \rangle$ is the local time variable within the interval and $\bullet_i$ abbreviates $\bullet(t_i)$ to render the notation compact.

Substituting \eqref{eq:vapprox} into \eqref{eq:rdvisco} reveals that the evolution of the restoring force $f_p$ satisfies
\begin{eqnarray}
\frac{\dot{f}_p(t_i+\tau)}{k_p} + \frac{f_p(t_i+\tau)}{\eta_p} = \dot{r}_i + \half(\ddot{r}_i + \ddot{r}_{i+1})\tau.
\label{eq:rdvisco_inc}
\end{eqnarray}
This Cauchy problem with the initial condition $f_p(t_i) = f_{p,i}$ has the solution 
\begin{align}
f_p(t_i+\tau) =
\nonumber \\
\left(f_{p,i} - \eta_p \dot{r}_i  + \frac{\eta_p^2}{2k_p} (\ddot{r}_i + \ddot{r}_{i+1}) \right ) \exp \left( -\frac{k_p}{\eta_p}\tau \right) \nonumber \\
+
\eta_p\dot{r}_i
+ \frac{\eta_p\tau k_p-\eta_p^2}{2k_p}
(\ddot{r}_i + \ddot{r}_{i+1}). 
\end{align}
Thus, at the end of the time interval with $\tau = \Delta t$, we have
\begin{align}\label{eq:fvfinal2} 
    f_{p,i+1} 
    =
    A_p f_{p,i} 
    +
    k_p 
    \widehat{\theta}_{p} 
    \dot{r}_i 
    + 
    B_p
    \left(\ddot{r}_i 
    +
    \ddot{r}_{i+1}
    \right),
\end{align}
where 
\begin{align}\label{eq:eff_relaxation}
\widehat{\theta}_{p} 
= 
\frac{\eta_p}{k_p} 
\left(
    1 - \exp \left( -\frac{k_p}{\eta_p}\Delta t \right)
\right)
\end{align}
denotes the effective relaxation time of the $p$-th cell and the auxiliary factors are given by
\begin{align}\label{eq:auxiliary_factors}
    A_p 
    & = 
    \left( 1 - \frac{k_p}{\eta_p} \widehat{\theta}_{p}\right),
    &
    B_p 
& =
\half \eta_p \left(\Delta t - 
\widehat{\theta}_{p}
\right).
\end{align}

Finally, substituting equations \eqref{eq:rapprox} and \eqref{eq:fvfinal2} into (\ref{eq:equilibrium}) expressed at $t_{i+1} = t_i +\Delta t$ and rearranging the terms yields
\begin{align}
& \Bigl( 
m + \quar k_{\infty} \Delta t^2 + 
\sum_{p=1}^P
B_p 
\Bigr)
\ddot{r}_{i+1}
= F_{i+1} - 
\sum_{p=1}^P A_p f_{p,i} 
\nonumber \\
- & 
k_{\infty} r_i 
- 
\Bigl(
k_{\infty}\Delta t +
\sum_{p=1}^P
k_p \widehat{\theta}_{p}
\Bigr)
\dot{r}_i 
\nonumber \\
- &
\Bigl(
\quar k_{\infty}\Delta t^2 
+ 
\sum_{p=1}^P B_p
\Bigr)
\ddot{r}_i.
\label{eq:eqof}
\end{align}

After solving~\Eref{eq:eqof} for the acceleration $\ddot{r}_{i+1}$, we update velocity $\dot{r}_{i+1}$, displacement $r_{i+1}$, and restoring forces $f_{p,i}$ according to equations \eqref{eq:vapprox}, \eqref{eq:rapprox}, and \eqref{eq:fvfinal2}, respectively, and proceed to the next time interval. 

Note that the initial acceleration $\ddot{r}_0$, needed in the first step of the algorithm, is set to
\begin{align}\label{eq:initial_acceleration}
\ddot{r}_0 = \frac{1}{m}
\bigl(
F(0) - k_\infty r_0 - \sum_{p=1}^Pf_{p,0}
\bigr),
\end{align}
according to the equilibrium~\eqref{eq:equilibrium} and initial~\eqref{eq:initial_conditions} conditions.   

\section{Variational integrators}\label{sec:vi}

Having derived the Newmark viscoelastic algorithm by conventional means, we now demonstrate its variational structure, by adapting the general arguments on variational integrators by Kane et~al.~\citep{Kane2000} to the current setting.

We will proceed in four steps. In \secref{sec:variational_framework}, we show that the governing equations from \secref{sec:governing} follow from the Euler-Lagrange~(E-L) equations of a suitably defined energy functional. Its discretization then provides the governing equations of the corresponding variational integrator introduced in \secref{sec:variational_discretization}. In \secref{sec:equivalence}, we demonstrate the equivalence of the integrator to the Newmark algorithm from \secref{sec:discretization}. In the last step, \secref{sec:energy_balance}, we comment on the energy conservation properties of the time integration scheme.

\subsection{Variational framework}\label{sec:variational_framework}
We postulate that the trajectory $\q : (0,T) \rightarrow Q$ of a constrained  dissipative mechanical system in the state space $Q$ is given by the Euler-Lagrange equations, e.g.,~\cite[Section~1.3]{Bedford1985}
\begin{subequations}
\label{eq:lagrangeEulerGen}
\begin{align} 
    \frac{\partial\mathcal{R}(\dot{\q}(t))}{\partial\dot{\q}}
    = & \, 
    \frac{%
        \partial \mathcal{L}(t, \q(t), \dot{\q}(t), \lam(t))
    }{\partial \q}
    \nonumber \\
    & -
    \frac{\partial}{\partial t}
    \frac{%
    \partial\mathcal{L}(t, \q(t), \dot{\q}(t), \lam(t))}{\partial\dot{\q}}, 
    \label{eq:ER_q} \\
    \mathbf{0} = & \, \bm{\alpha}(\q(t)).
    \label{eq:ER_lam}
\end{align}
\end{subequations}    
Here, $\mathcal{R}$ stands for the dissipation potential, $\mathcal{L}$ for the Lagrangian of the problem, and $\lam : (0,T) \rightarrow \Lambda$ denotes the Lagrange multipliers associated with the kinematic constraint function $\bm{\alpha}$. Besides, these equations correspond to the stationarity conditions of the action functional 
\begin{align}
    \mathcal{S}( \trial\q, \trial\lam )
    =
    \int_0^T
        \mathcal{L}
        \bigl(
            t, \trial{\q}(t), \trial{\dot{\q}}(t), \trial\lam(t)
        \bigr)
    \de t
\label{eq:actionIntegral}
\end{align}
perturbed by the dissipative forces $\partial_{\dot \q} \mathcal{R}$. Note that the hat symbol in~\eqref{eq:actionIntegral} now distinguishes the test quantities from the true trajectories defined with~\eqref{eq:lagrangeEulerGen}.

For the problem from \figref{fig:maxwellP}, the state variable 
\begin{align}
    \trial{\q}(t)
    =
    \begin{bmatrix}
         \trial{r}(t), \,
        \{ \trial{r}(t)_{\mathrm{e},p}\}_{p=1}^P, \, 
        \{ \trial{r}(t)_{\mathrm{v},p}\}_{p=1}^P  \,
    \end{bmatrix}\trn
\end{align}
collects the total displacement and the displacements of both components of each Maxwell cell; the state space $Q = \mathbb{R}^{2P+1}$. The Lagrangian has the standard form 
\begin{align}
    \mathcal{L}(t,\trial\q, \trial{\dot{\q}}, \trial\lam)
    = & \;
    \mathcal{K}(\trial{\dot{\q}})
    -
    \mathcal{E}(\trial\q)
    +
    \bm{f}_{\mathrm{ext}}(t)\trn \trial\q
    \nonumber \\
    &+
    \trial\lam\trn\bm{\alpha}(\trial\q)
    \label{eq:lagrangian}
\end{align}    
involving the kinetic energy $\mathcal{K}$, potential energy of deformation $\mathcal{E}$, and external forces $\bm{f}_{\mathrm{ext}}$ given by
\begin{subequations}
    \begin{align}
        \mathcal{K}(\trial{\dot \q}) 
        & = \half m ( \trial{\dot{r}} )^2, \\
        \mathcal{E}(\trial\q)
        & 
        = 
        \half k_\infty \trial{r}^2
        + 
        \half
        \sum_{p=1}^P
        k_p (\trial{r}_{\mathrm{e},p})^2, 
        \\
        \bm{f}_\mathrm{ext}( t ) 
        & = 
        \begin{bmatrix}
        F(t), \, \bm{0}_{1 \times 2P}
        \end{bmatrix}\trn.
    \end{align}
\end{subequations}
The kinematical constraints take the form
\begin{align}
    \alpha_p( \trial{\q} )
    = 
    \trial{r}_{\mathrm{e},p} 
    + 
    \trial{r}_{\mathrm{v},p}
    -
    \trial{r},
    && 
    p = 1, \ldots, P;        
\end{align}
the space of the Lagrange multiplies $\Lambda$ then becomes $\mathbb{R}^P$. The last component of the general framework~\eqref{eq:lagrangeEulerGen} is provided by the dissipation potential
\begin{align}
    \mathcal{R}(\trial{\dot{\q}}) 
    = 
    \half
    \sum_{p=1}^{P}
    \eta_p ( \trial{\dot{r}}_{\mathrm{v},p} )^2    
\end{align}
involving solely the viscous displacements of all cells.

In this setting, the E-L equation~\eqref{eq:ER_q} represents the system of $(1 + 2P)$ optimality conditions. The first one, corresponding to the total displacement $r$, attains the form
\begin{align}
    m\ddot{r}(t)
    +
    k_\infty r(t)
    +
    \sum_{p=1}^P\lambda_p( t ) 
    = 
    F(t),
    \label{eq:varequilibrium}
\end{align}
while the remaining $2P$ conditions read as
\begin{align}\label{eq:varLambda}
\lambda_p( t )
=
k_p r_{\mathrm{e},p}( t ), 
&& 
\lambda_p( t )
=
\eta_p
\dot{r}_{\mathrm{v},p} ( t ),
\end{align}    
with $p = 1, \ldots, P$. It is thus evident that the multipliers $\lambda_p$ play role of the viscous force $f_p$ and, because the optimality~\eqref{eq:ER_lam} and compatibility~\eqref{eq:rdvisco} conditions coincide, the current setting is equivalent to the one of \secref{sec:governing}.

\subsection{Discretization}\label{sec:variational_discretization}

Recall that the incremental algorithm of \secref{sec:discretization} relies on the discretization of the total displacements, from which the evolution of cell-related variables $r_{\mathrm{e}, p}$, $r_{\mathrm{v}, p}$, and $f_p$ follows in the closed form. To mimic this structure, only the total displacements $r$ will be determined from the discrete (non-dissipative) E-L equations, whereas the remaining quantities are determined from the non-discretized optimality conditions~\eqref{eq:varLambda} and~\eqref{eq:ER_lam}.   

To this goal, we consider the same discretization of the time interval $\langle 0, T \rangle$ as in \secref{sec:discretization} and introduce the discretized action functional 
\begin{align}
    \mathcal{S}(
        \trial{r}, 
        \trial{\lam}
        )
    & \approx
    \mathcal{S}\discr
    \bigl( 
    \{ \trial{r}_i \}_{i=0}^N, 
    \{ \trial{\lam}_i\}_{i=0}^N 
    \bigr)
    \nonumber \\ 
    & =
    \Delta t
    \sum_{i=0}^{N-1}
     \mathcal{L}\discr \bigl(
         \trial{r}_i, 
         \trial{r}_{i+1}, 
         \trial{\lam}_i, 
         \trial{\lam}_{i+1} \bigr),
    \label{eq:actionIntegralAprx}
\end{align}
with the discrete Lagrangian given by~\cite[Eq.~(2)]{Kane2000}
\begin{align}
    \mathcal{L}\discr \bigl(
        \trial{r}_i, 
        \trial{r}_{i+1}, 
        \trial{\lam}_i, 
        \trial{\lam}_{i+1} 
        \bigl)
    & = 
    \half m \bigl( 
        \frac{\trial{r}_{i+1} 
        - 
        \trial{r}_i}{\Delta t} \bigr)^2 
    \\ 
    & 
    -
    \half k_\infty \bigl( 
        \frac{\trial{r}_{i+1} 
        + 
        \trial{r}_i}{2} 
        \bigr)^2
    \nonumber \\
    & +
    \quar ( \trial{r}_i  + \trial{r}_{i+1} ) (F_i + F_{i+1} )
    \nonumber \\ 
    & \hspace{-0.5em} 
    -
    \quar 
    \sum_{p=1}^P
    ( \trial{r}_i  + \trial{r}_{i+1} ) 
    ( \trial{\lambda}_{p,i} + \trial{\lambda}_{p,i+1} ).
    \nonumber
\end{align}

The stationarity conditions at time $t_i$, $\partial \mathcal{S}\discr / \partial r_i = 0$ with $i = 1, \ldots, N-1$ read as
\begin{align}\label{eq:DEL}
    0 & = 
    \frac{
        \partial\mathcal{L}\discr(r_{i-1}, r_i, \lam_{i-1}, \lam_{i})
    }{\partial r_i}
    \nonumber \\
    & + 
    \frac{\partial\mathcal{L}\discr(r_i, r_{i+1}, \lam_i, \lam_{i+1})
    }{\partial r_i}
\end{align}
which delivers the governing equations of the variational integrator in the form\footnote{%
Notice that we assume the Lagrange multipliers $\lambda_{p,i}$ to be given abritrary quantities, similarly to the forcing terms $F_i$. Once we establish the equivalance to the Newmark algorithm, their values follow from the update formula~\eqref{eq:fvfinal2} from Section~\ref{sec:Newmark}.}
\begin{align}\label{eq:discreteeqv}
    & m
    \frac{r_{i+1}-2r_{i}+r_{i-1}}{\Delta t^2}
    +
    \quar
    k_\infty
    ( r_{i+1}+2r_{i}+r_{i-1} )
\nonumber \\
    + &  \;
    \sum_{p=1}^P
    \quar
    ( \lambda_{p,i+1} + 2 \lambda_{p,i} + \lambda_{p,i-1} )
\nonumber \\
    = & \;
    \quar
    ( F_{i-1}+2F_{i}+F_{i+1} ).
\end{align}

\subsection{Equivalence to Newmark}\label{sec:equivalence}

We will proceed with additional two steps to show that the optimality conditions~\eqref{eq:discreteeqv} correspond to the Newmark integration scheme from \secref{sec:Newmark}. First, we demonstrate that the displacements $\{r_i\}_{i=0}^N$ provide definitions of velocities~$\{\dot{r}_i\}_{i=0}^N$ and accelerations $\{\ddot{r}_i\}_{i=0}^N$ consistent with the kinematic assumptions in Eq.~\eqref{eq:Newmark_kinematics}. Second, we show that the discrete-in-time quantities satisfy the equations of motion~\eqref{eq:equilibrium}. 

\paragraph{Kinematics.} Following~\cite[Section~2.2]{Kane2000}, we start from introducing auxiliary accelerations  
\begin{align}
    m\ddot{r}_{i+1/2} & =
    -
    \frac{k_\infty}{2}(r_i+r_{i+1})
    +
    \half(F_i+F_{i+1})
    \nonumber \\
    & -
    \sum_{p=1}^P
    \half( \lambda_{p,i} + \lambda_{p,i+1}), 
    \label{eq:acc_plus}
\end{align}
for $i=0, 1, \ldots, N-1$. Summing $m \ddot{r}_{i-1/2}$ with $m \ddot{r}_{i+1/2}$ comparing the result with~\eqref{eq:discreteeqv} provides
\begin{align}\label{eq:riigoverning}
    \frac{
        r_{i+1}-2r_{i}+r_{i-1}
        }{
            \Delta t^2
        }
    =
    \half( \ddot{r}_{i+1/2}+\ddot{r}_{i-1/2} ),
\end{align}
with $i=1, \ldots, N-1$. 

The discrete linear momenta follow standardly from
\begin{align}\label{eq:linear_momenta_def}
    p_i 
    =
\frac{
    \partial\mathcal{L}\discr(r_{i-1}, r_{i}, \lam_{i-1}, \lam_i)
    }{
\partial r_i}
\Delta t,
\end{align} 
and, using $\ddot{r}_{i-1/2}$ from Eq.~\eqref{eq:acc_plus}, they can be evaluated as
\begin{align}\label{eq:linear_momenta_i}
    p_i 
    =
    m \dot{r}_i
    =
    m\frac{r_i-r_{i-1}}{\Delta t}
    +
    m \frac{\ddot{r}_{i-1/2}}{2}
    \Delta t.
\end{align}
Expressing $p_{i+1}$ according to the previous relation and employing~\eqref{eq:riigoverning} provides 
\begin{align}
    p_{i+1} 
    =
    p_i 
    +
    m \ddot{r}_{i+1/2} \Delta t,
\end{align} 
from which we obtain
\begin{align}\label{eq:veloc1}
    \dot{r}_{i+1} 
    = 
    \dot{r}_{i} 
    + 
    \Delta t 
    \ddot{r}_{i+1/2}. 
\end{align}

Likewise, expressing $r_i$ from~\eqref{eq:riigoverning} and employing the velocity $\dot{r}_i$ from~\eqref{eq:linear_momenta_i} provides
\begin{align}\label{eq:displ1}
    r_{i+1}
    =
    r_{i} 
    +
    \dot{r}_{i} \Delta t
    +
    \half
    \ddot{r}_{i+1/2} \Delta t^2.
\end{align}
Hence, expressions~\eqref{eq:displ1} and~\eqref{eq:veloc1} become identical to the ones of the Newmark method~\eqref{eq:Newmark_kinematics} once setting
\begin{align}\label{eq:nodal_acceration_vi}
    \ddot{r}_{i+1/2} = \half( \ddot{r}_i + \ddot{r}_{i+1} ).
\end{align}

\paragraph{Equilibrium.} Employing the nodal accelerations $\ddot{r}_{i-1/2}$ and $\ddot{r}_{i+1/2}$ from~\eqref{eq:nodal_acceration_vi} in the identity~\eqref{eq:riigoverning} reveals that
\begin{align}\label{eq:equiv_acceleration}
    \frac{r_{i+1}-2r_{i}+r_{i-1}}{\Delta t^2}
    =
    \quar
    ( \ddot{r}_{i-1} + 2 \ddot{r}_{i} + \ddot{r}_{i+1} ).
\end{align}
Further, by expressing the difference $m(\ddot{r}_{i+1/2} - \ddot{r}_{i-1/2})$ using~\eqref{eq:acc_plus}, we find that 
\begin{align}\label{eq:accelerations_diff}
  &  \half m ( \ddot{r}_{i+1} - \ddot{r}_{i-1} )
    +
    \half k_\infty (r_{i+1} - r_{i-1})
    \nonumber \\
+ \; &
    \sum_{p=1}^P
    \half( \lambda_{p,{i+1}} - \lambda_{p,i-1}) 
    = 
    \half(F_{i+1}-F_{i-1}).
\end{align}
Now, after inserting the identity~\eqref{eq:equiv_acceleration} into the discrete Euler-Lagrange equations~\eqref{eq:discreteeqv} and subtracting~\eqref{eq:accelerations_diff} from the result, we infer that
\begin{align}\label{eq:diff_final}
    & m ( \ddot{r}_{i-1} + \ddot{r}_i )
    +
    k_\infty ( r_{i-1} + r_i )
    +
    \sum_{p=1}^P
    ( \lambda_{p,{i-1}} + \lambda_{p,i} )
    \nonumber \\
    = &
    F_{i-1} + F_i,
\end{align}
which can be reduced to the final form
\begin{align}\label{eq:equilibrium_eqs}
    m \ddot{r}_i 
    +
    k_\infty r_{i}
    +
    \sum_{p=1}^P
    \lambda_{p,i} 
    =
    F_i.
\end{align}
Indeed, the equivalence between~\eqref{eq:equilibrium_eqs} and~\eqref{eq:diff_final} for $i=1$ holds because of the choice of the initial acceleration~\eqref{eq:initial_acceleration}, and for $i = 2, \ldots, N-1$ it follows by induction. 

\subsection{Energy balance}\label{sec:energy_balance}

The variational framework~\eqref{eq:lagrangeEulerGen} additionally reveals that the trajectory $\q$ satisfies the energy balance condition, e.g.,~\cite[Section~5.1]{mielke2015rate}
\begin{align}\label{eq:energy_equality}
\mathcal{E}_\mathrm{int}(t)
+
\mathcal{D}(t)
=
\mathcal{E}_\mathrm{int}(0)
+
\mathcal{W}(t)
& 
\text{ for } 0 \leq t \leq T, 
\end{align}
with the internal energy $\mathcal{E}_\mathrm{int}$, dissipated energy $\mathcal{D}$, and the work done by external forces $\mathcal{W}$ given by     
\begin{subequations}
\begin{align}
\mathcal{E}_\mathrm{int}(t)
& =
\half m\dot{r}^2(t)
+
\half k_\infty r^2(t)
\nonumber \\
& + 
\sum_{p=1}^P\half k_pr_{\mathrm{e},p}^2(t),
\label{eq:internalEnergy}
\\
\mathcal{D}(t)
& =
\sum_{p=1}^P
\int_0^t\eta_p\dot{r}_{\mathrm{v},p}^2(\tau)
\mathrm{d}\tau,
\label{eq:dissipationEnergy}
\\
\mathcal{W}(t)
& = 
\int_0^tF(\tau)\dot{r}(\tau)\mathrm{d}\tau.
\label{eq:externalEnergy}
\end{align}
\label{eq:enegy_balance}
\end{subequations}

To later quantify the articifial dissipation induced by time discretization, we also consider the time-discrete quantities 
\begin{subequations}
\begin{align}
\mathcal{E}_\mathrm{int}(t_i)
& =
\half m\dot{r}^2_{i}
+
\half k_\infty r_i^2
+
\sum_{p=1}^P
\frac{\lambda_{p,i}^2}{2k_p},
\label{eq:internalEnergy_discr}
\\
\mathcal{D}\discr(t_i)
& =
\sum_{k=0}^{i-1}
\sum_{p=1}^P
\frac{1}{2\eta_p}\bigl(
    \lambda_{p,k}^2
    +
    \lambda_{p,k+1}^2 
\bigr)
\Delta t,
\label{eq:dissipationEnergy_discr}
 \\
\mathcal{W}\discr(t_i) 
& =
\sum_{k=0}^{i-1}
\half \bigl(F_k\dot{r}_k+F_{k+1}\dot{r}_{k+1} \bigr)
\Delta t;
\end{align}
\label{eq:discretized_energies}%
\end{subequations}
the last two expressions correspond to the approximations of integrals in~\eqref{eq:enegy_balance} with the trapezoidal rule and employing the indentities~\eqref{eq:varLambda}.

\section{Examples}

In this section, we demonstrate the performance of the developed Newmark algorithm with two examples. The first one in \secref{sec:example_discrete_problem} addresses the accuracy and numerical energy dissipation of the integrator for the single-degree-of-freedom system from \figref{fig:maxwellP}. The follow-up example in \secref{sec:example_continuum} outlines an extension of the scheme towards continuum models. 

\begin{table}[ht]
    \centering
    \begin{tabular}{cc|cc}
    $k_p$ [kNm$^{-1}$] & $\theta_p$ [s] & $k_p$ [kNm$^{-1}$] & $\theta_p$ [s] \\
    \hline
    6933.9 & 10$^{-9}$ & 445.1 & 10$^{2}$ \\
    3898.6 & 10$^{-8}$ & 300.1 & 10$^{3}$ \\
    2289.2 & 10$^{-7}$ & 401.60 & 10$^{4}$ \\
    1672.7 & 10$^{-6}$ & 348.1 & 10$^{5}$ \\
    761.60 & 10$^{-5}$ & 111.6 & 10$^{6}$ \\
    2401.0 & 10$^{-4}$ & 127.2 & 10$^{7}$ \\
    65.200 & 10$^{-3}$ & 137.8 & 10$^{8}$ \\
    248.00 & 10$^{-2}$ & 50.5 & 10$^{9}$ \\
    575.60 & 10$^{-1}$ & 322.9 & 10$^{10}$ \\
    56.30 & 10$^{0}$ & 100.0 & 10$^{11}$ \\
    188.6 & 10$^{1}$ & 199.9 & 10$^{12}$ \\

    \end{tabular}
    \caption{Parameters of Maxwell chain model~\cite{Hana2019}, with $\theta_p=\eta_p/k_p$ and $k_\infty=682.18$ kNm$^{-1}$. Note that in \secref{sec:example_continuum}, the stiffnesses $k_\bullet$ correspond to shear moduli $G_\bullet$~[MPa].}
    \label{tab:pronySeries}
\end{table}
 
Data of the generalized Maxwell chain used in both examples appear in \tabref{tab:pronySeries}; they represent real PVB material with sufficiently short and long relaxation times for testing algorithm robustness. For more information on experimental procedures to determine these parameters, see~\cite{Hana2019}. 
All results presented in this section are reproducible with Python-based scripts available at~\cite{schdmit_jaroslav_2019_3265058}.

\subsection{Discrete problem}\label{sec:example_discrete_problem}
We consider the following two types of loading:
\begin{subequations} 
\label{eq:loading}
\begin{align}
    F(t) & = \overline{F} & \text{for } t \geq 0, \label{eq:ramp_load}\\ 
    F(t) & = \overline{F} \sin t & \text{for } t \geq 0,
\end{align}
\end{subequations}
corresponding to ramp and harmonic loads, respectively. In both cases, we set the amplitude $\overline F = 1$~MN and the mass $m=10^6$~kg to scale the displacement amplitude to $\approx 1$~m. Initial displacement, velocity, and forces in Maxwell cells were set to zero; recall~\eqref{eq:initial_conditions}. As for the Newmark algorithm, we set the time steps $\Delta t$ to $1.0$, $0.5$, and $0.2$~s. 

\paragraph{Accuracy} of the Newmark algorithm is checked by comparing its trajectories with the reference ones, obtained with the adaptive solver \verb!lsoda!~\cite{petzold_automatic_1983} --- available through \verb!odeint! function of Scipy library~\cite{Scipy} --- applied to the full initial value problem~\eqref{eq:equilibrium}, \eqref{eq:rdvisco}, and~\eqref{eq:initial_conditions}.

\begin{figure*}[h]
    \centering
    \begin{tabular}{cc}
    \includegraphics[width=.465\textwidth]{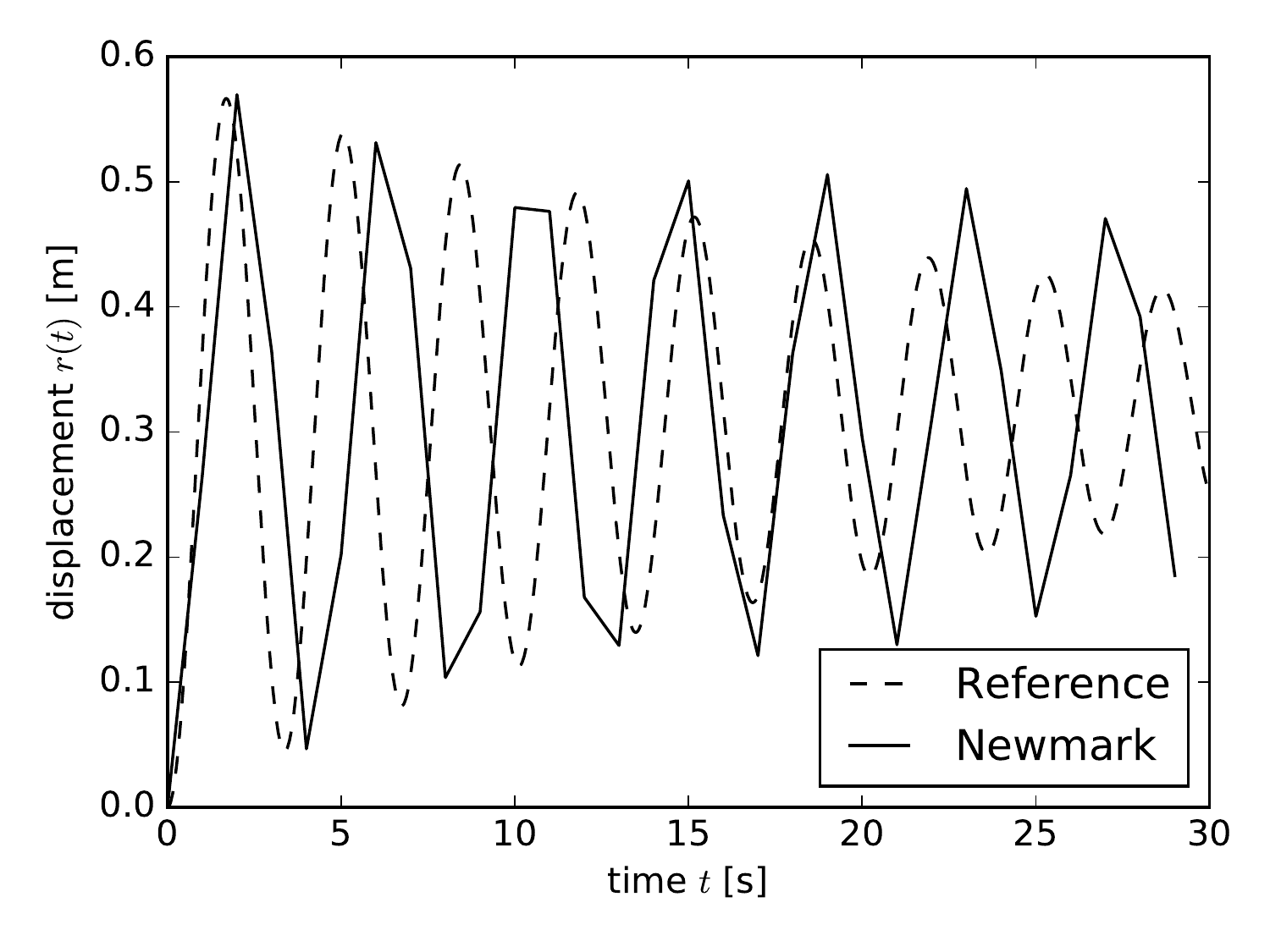} &
    \includegraphics[width=.465\textwidth]{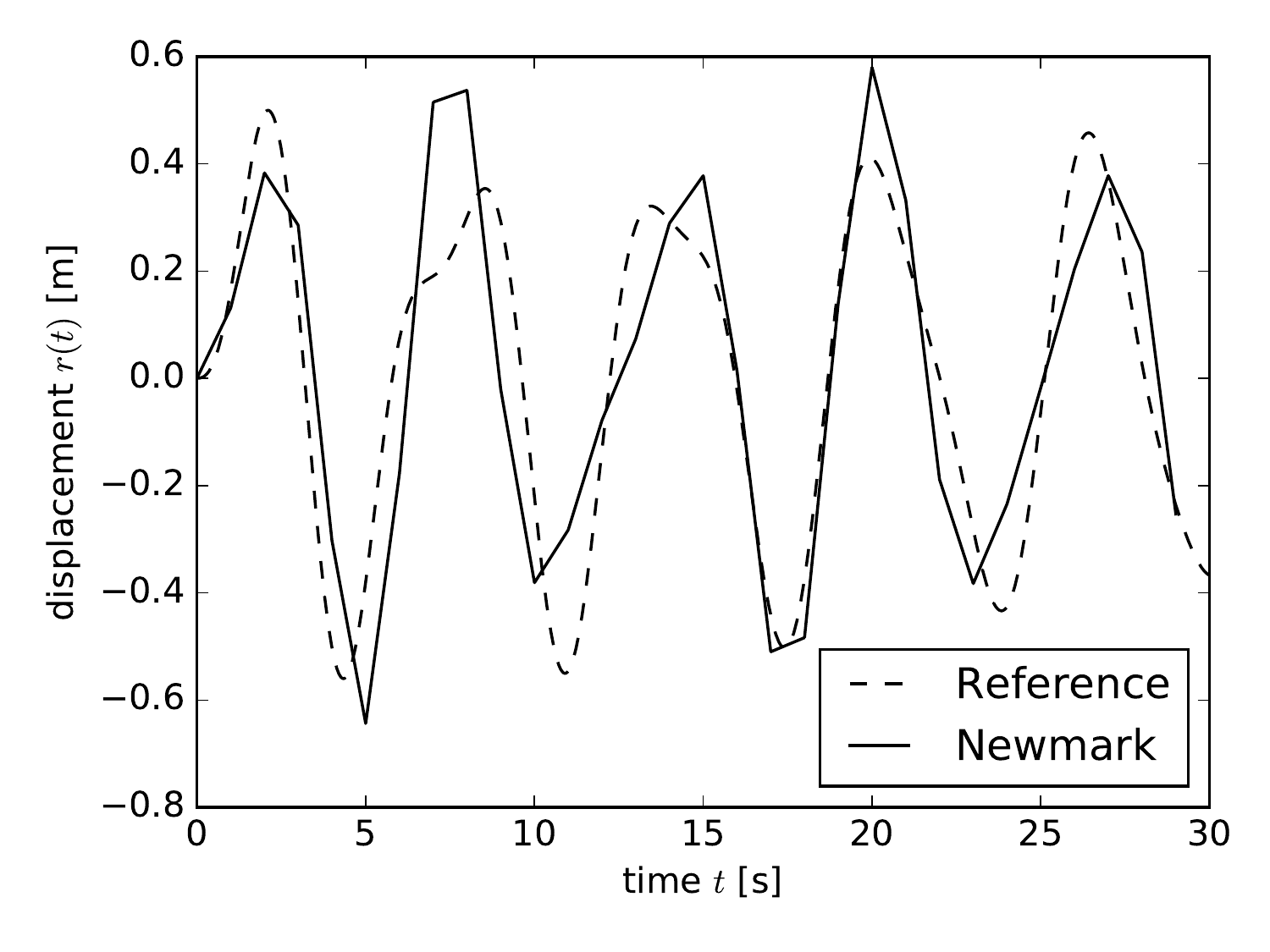} \\
    \includegraphics[width=.465\textwidth]{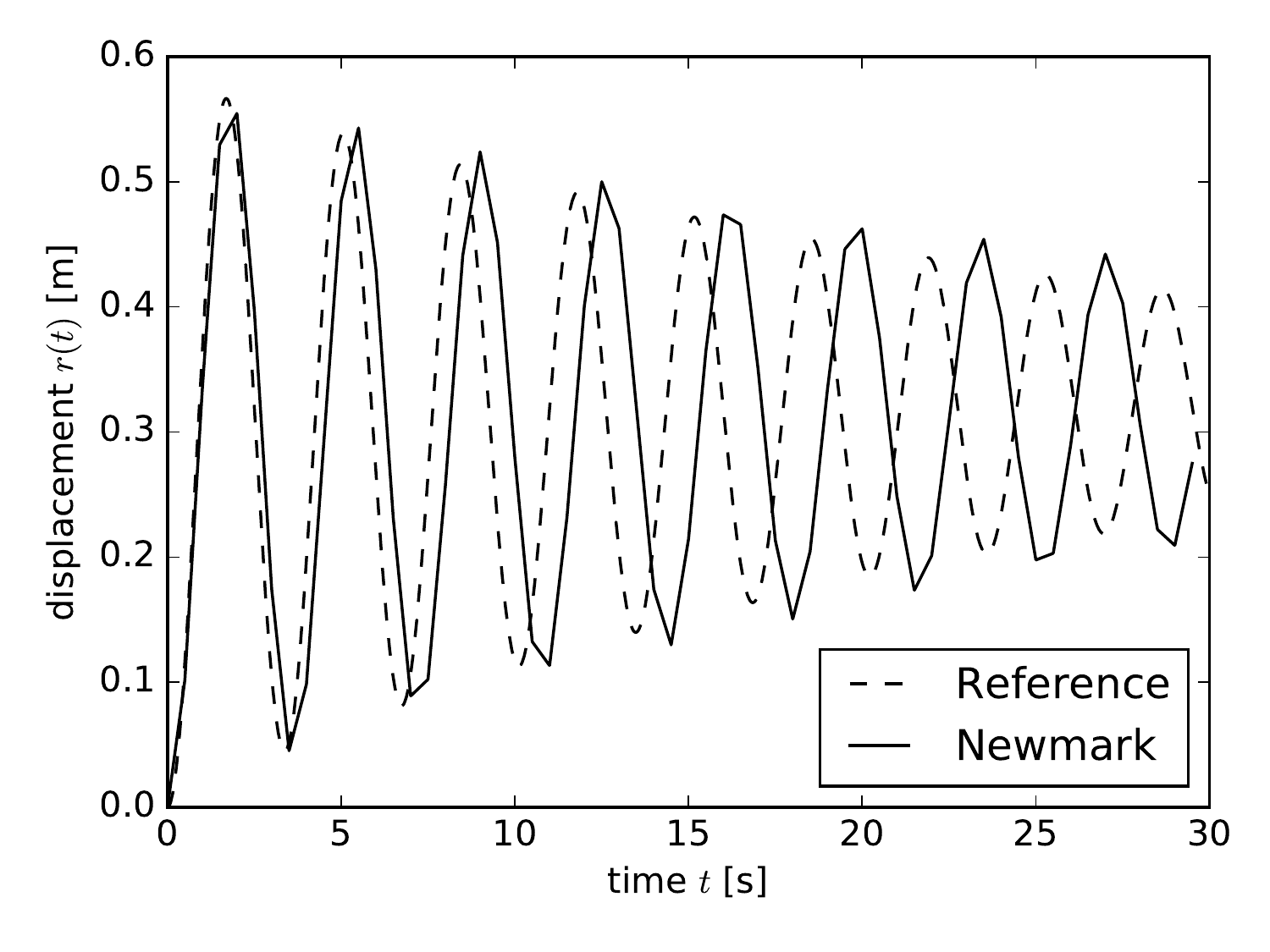} &
    \includegraphics[width=.465\textwidth]{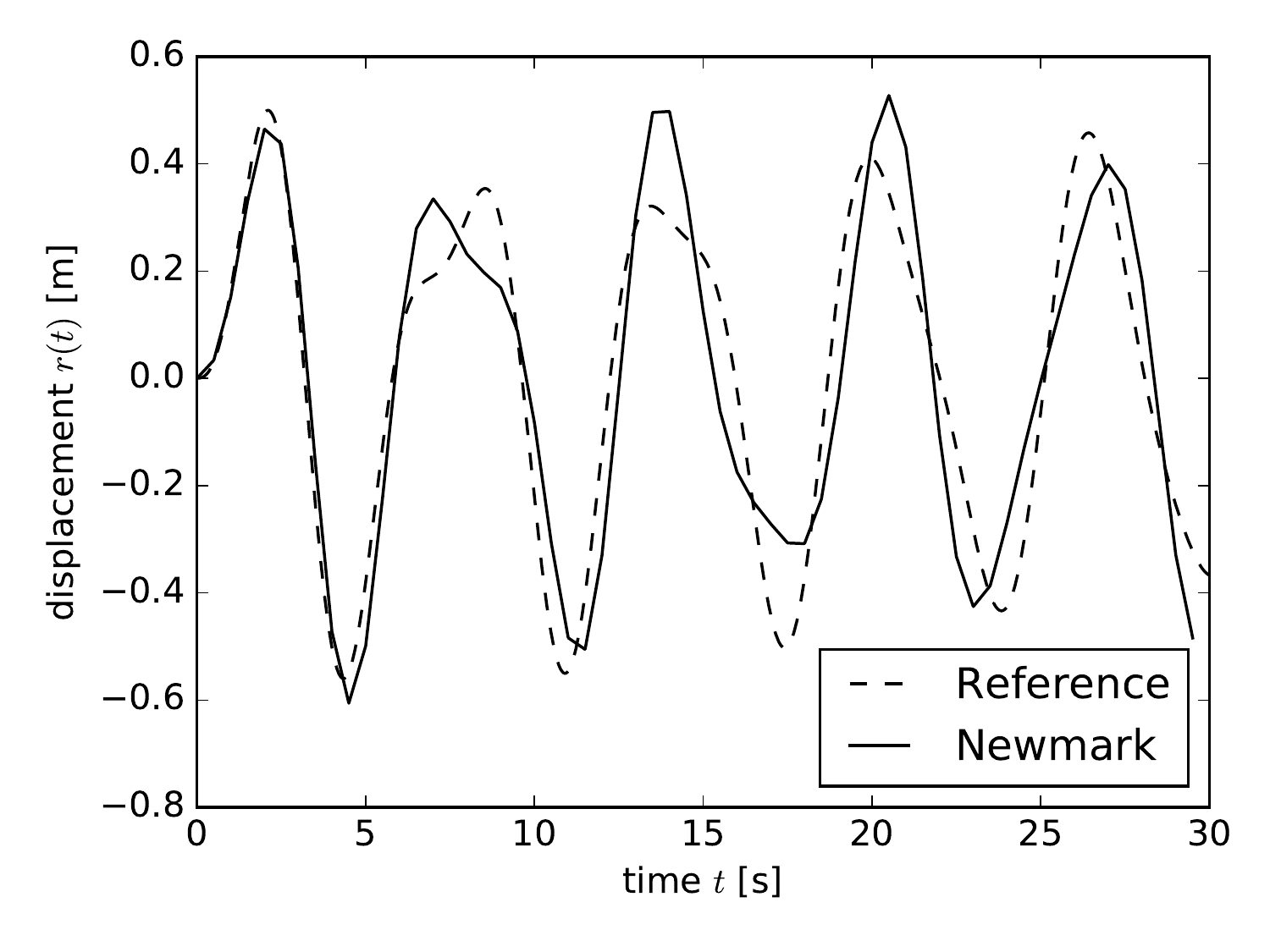} \\
    \includegraphics[width=.465\textwidth]{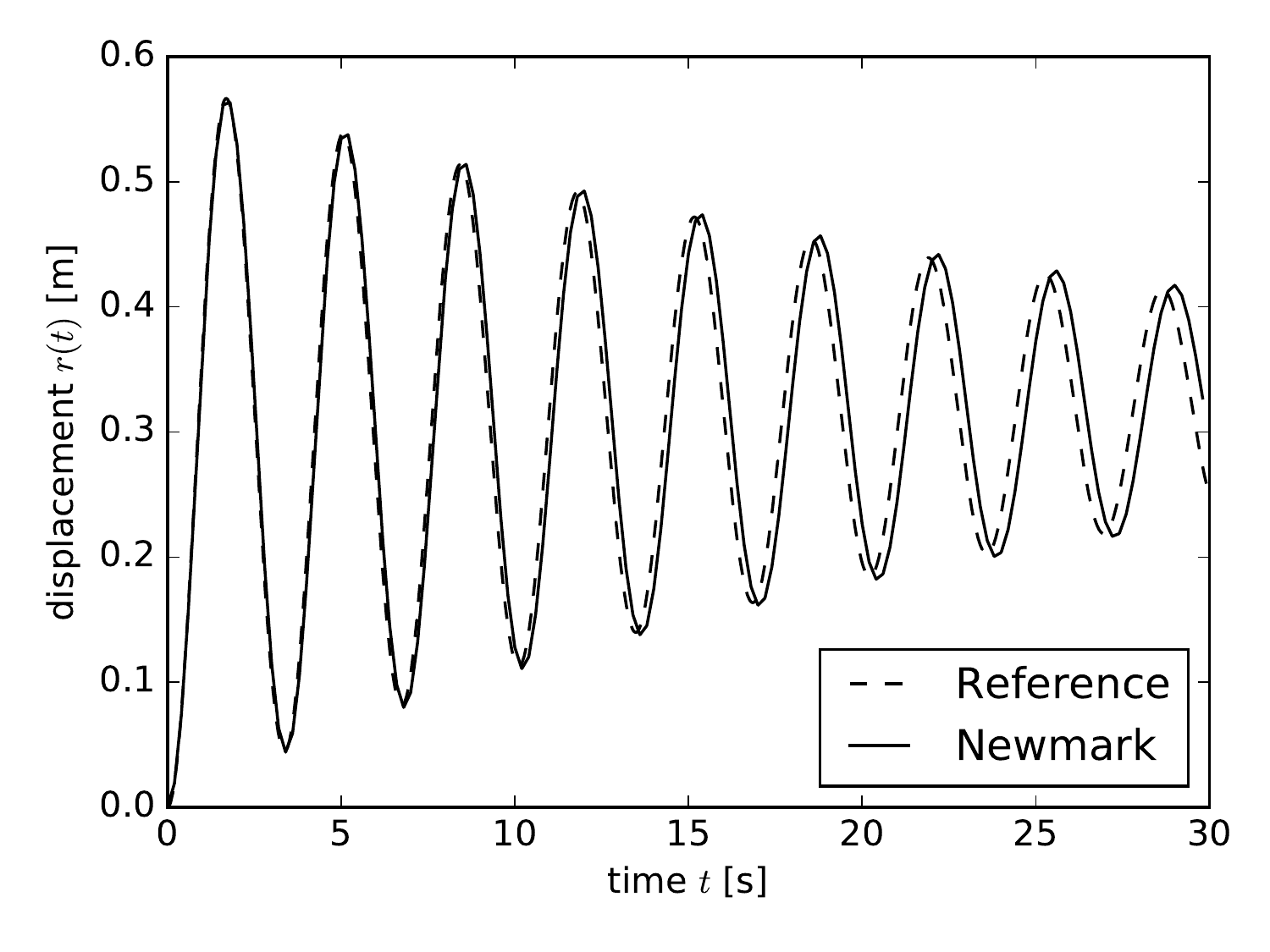} &
    \includegraphics[width=.465\textwidth]{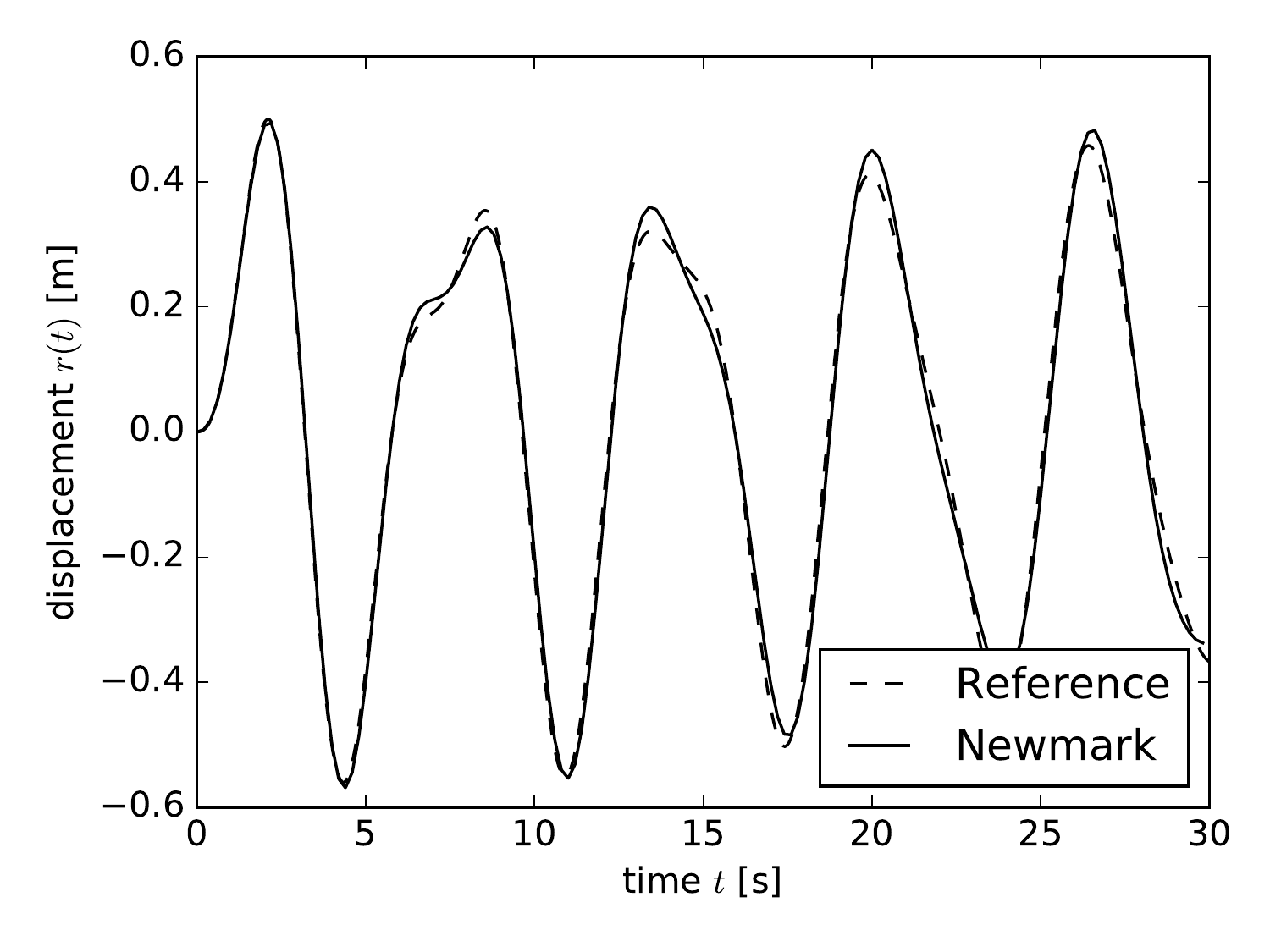} 
    \end{tabular}
    \caption{Accuracy of the viscous Newmark method for step~(left) and harmonic~(right) loadings defined by Eq.~\eqref{eq:loading}. Top, center, and bottom graphs show trajectories for time steps $\Delta t=1.0$~s, $\Delta t=0.5$~s, and $\Delta t=0.2$~s respectively.}
    \label{fig:sdof_results}
\end{figure*}

Results appear in \figref{fig:sdof_results} and demonstrate that the Newmark algorithm is stable even for coarse time steps, thanks to its variational structure. The errors behave consistently with findings for Newmark-family methods applied to linearly dampened systems, e.g.~\cite[Section~B.II.5]{Hughes1983}. In particular, the numerical dispersion (understood as the error in periods) and dissipation (error in amplitudes) decays as $\mathcal{O}( (\omega \Delta t)^2 )$, where $\omega$ stands for the angular frequency of the response. For $\Delta t = 0.2$~s, the trajectories predicted by the Newmark scheme closely match the reference ones. 

\begin{figure*}[h]
    \centering
    \begin{tabular}{cc}
    \includegraphics[width=.465\textwidth]{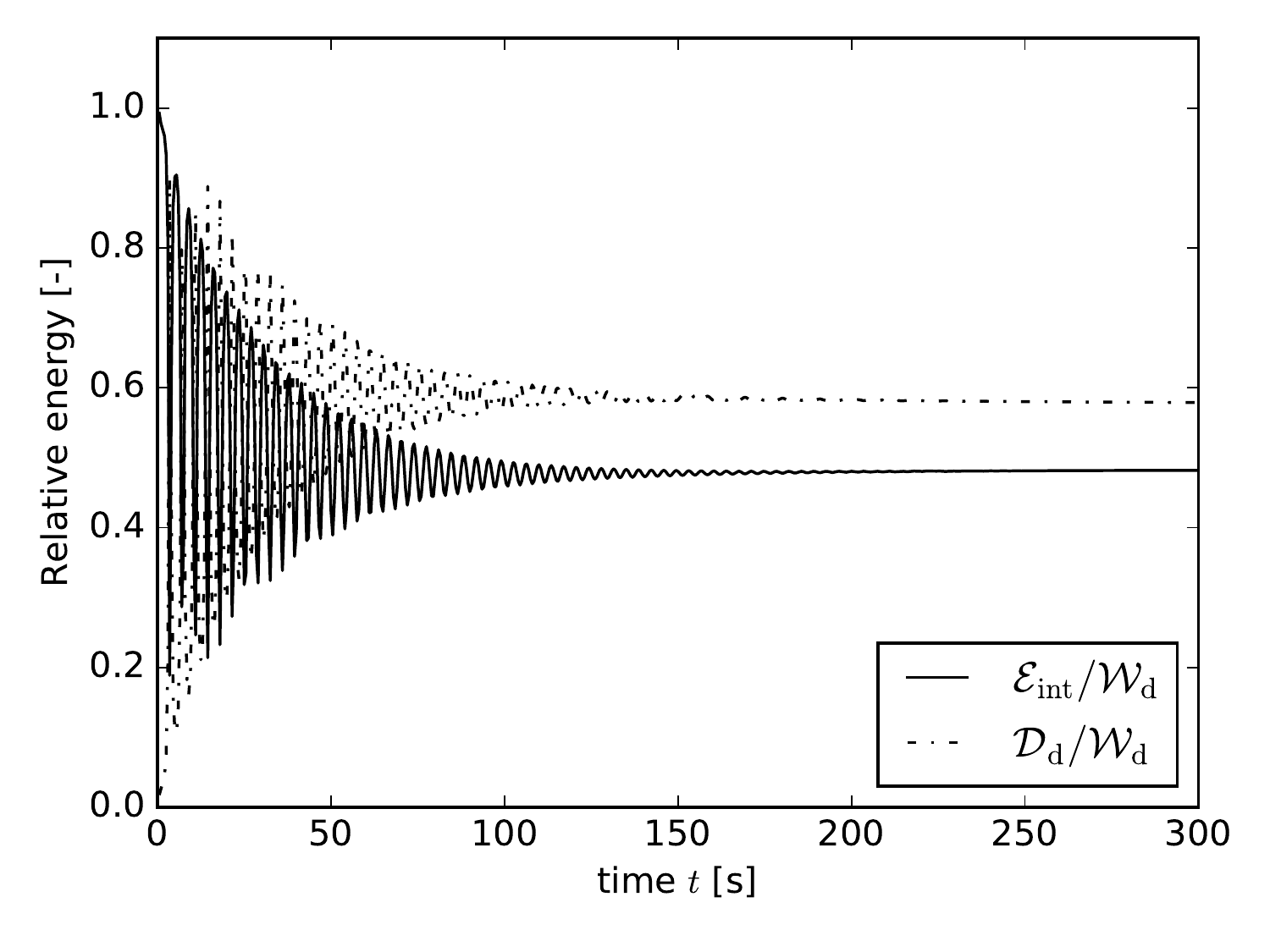} &   
    \includegraphics[width=.465\textwidth]{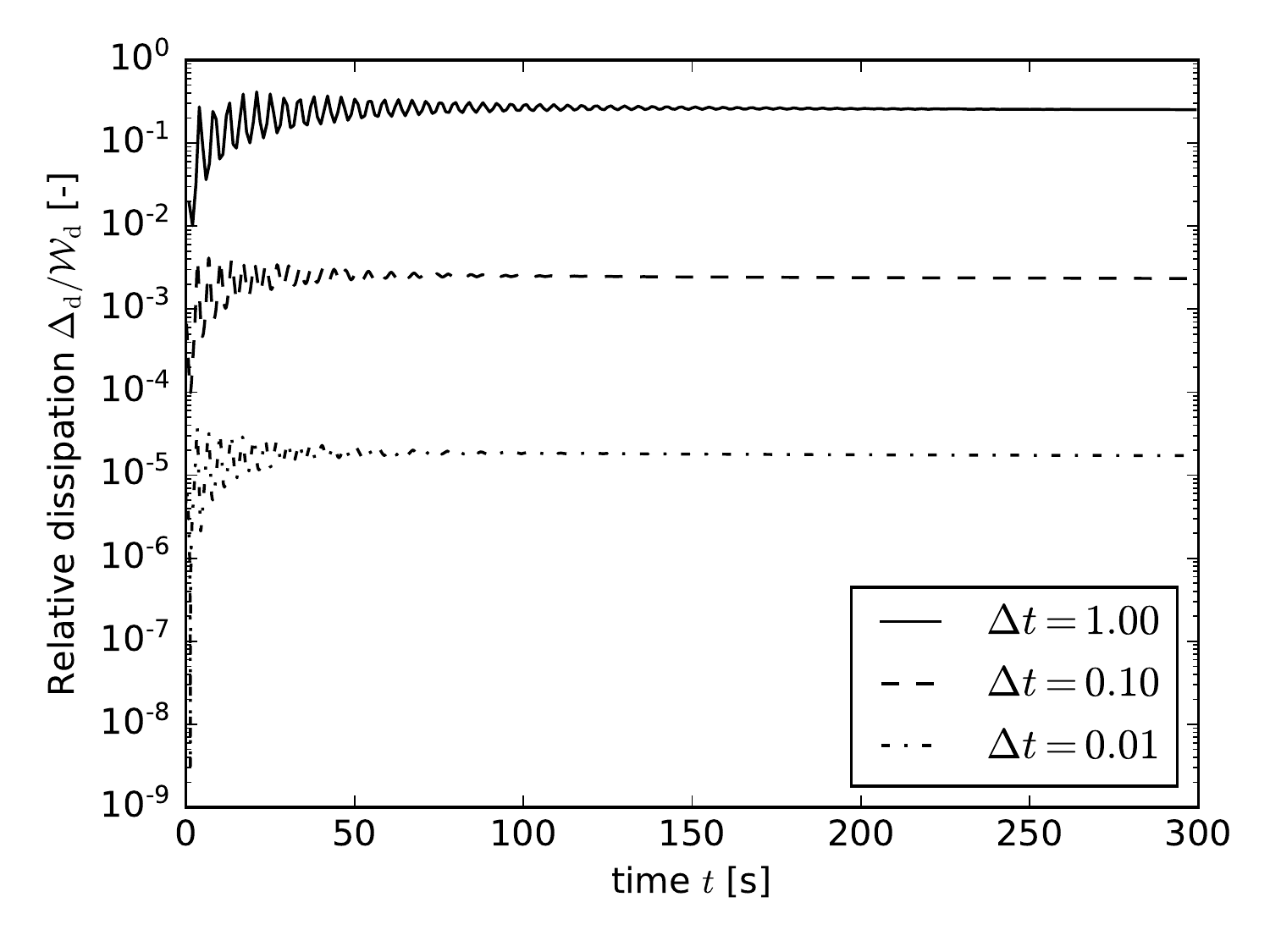} \\
    \includegraphics[width=.465\textwidth]{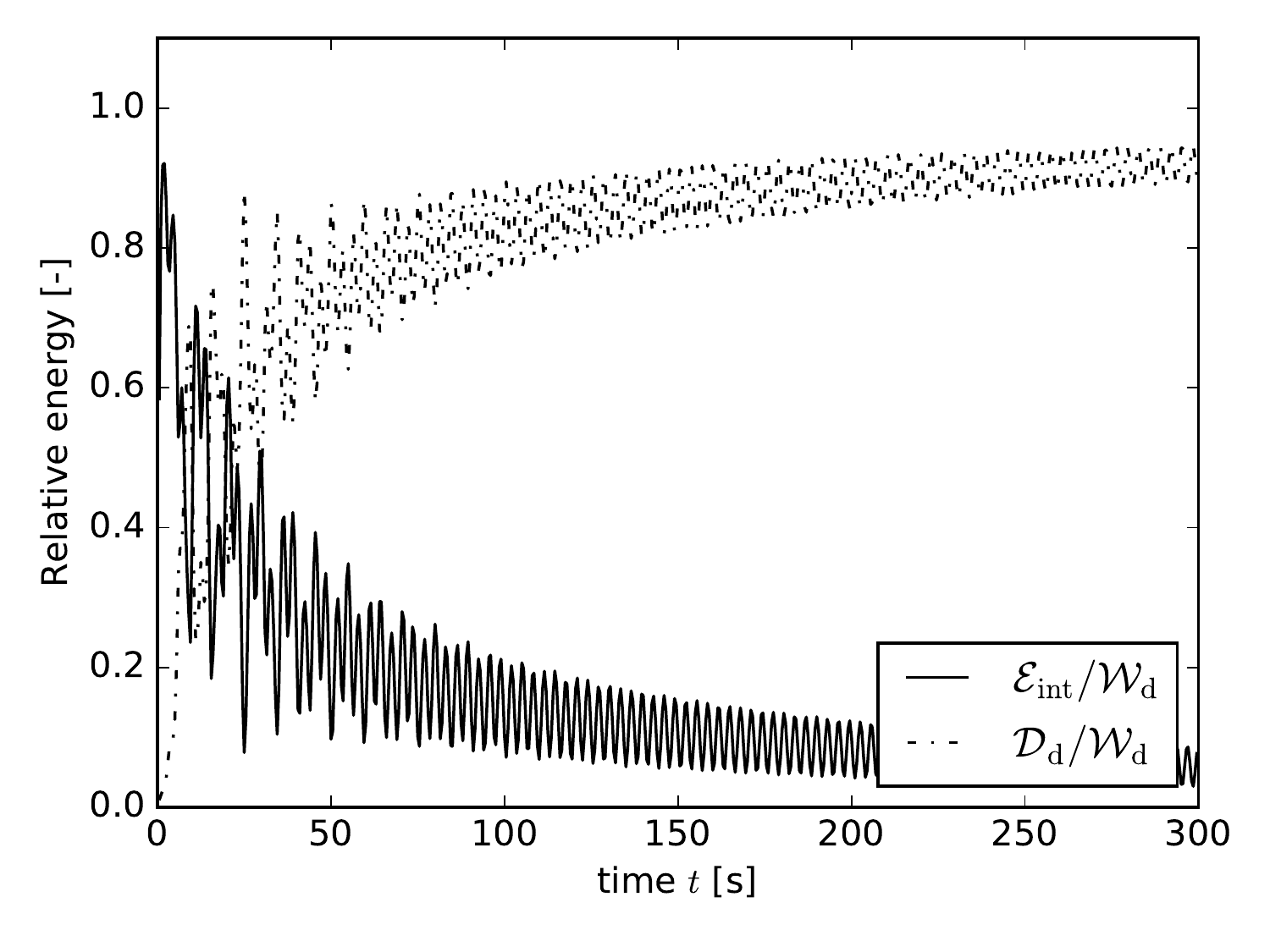} & 
    \includegraphics[width=.465\textwidth]{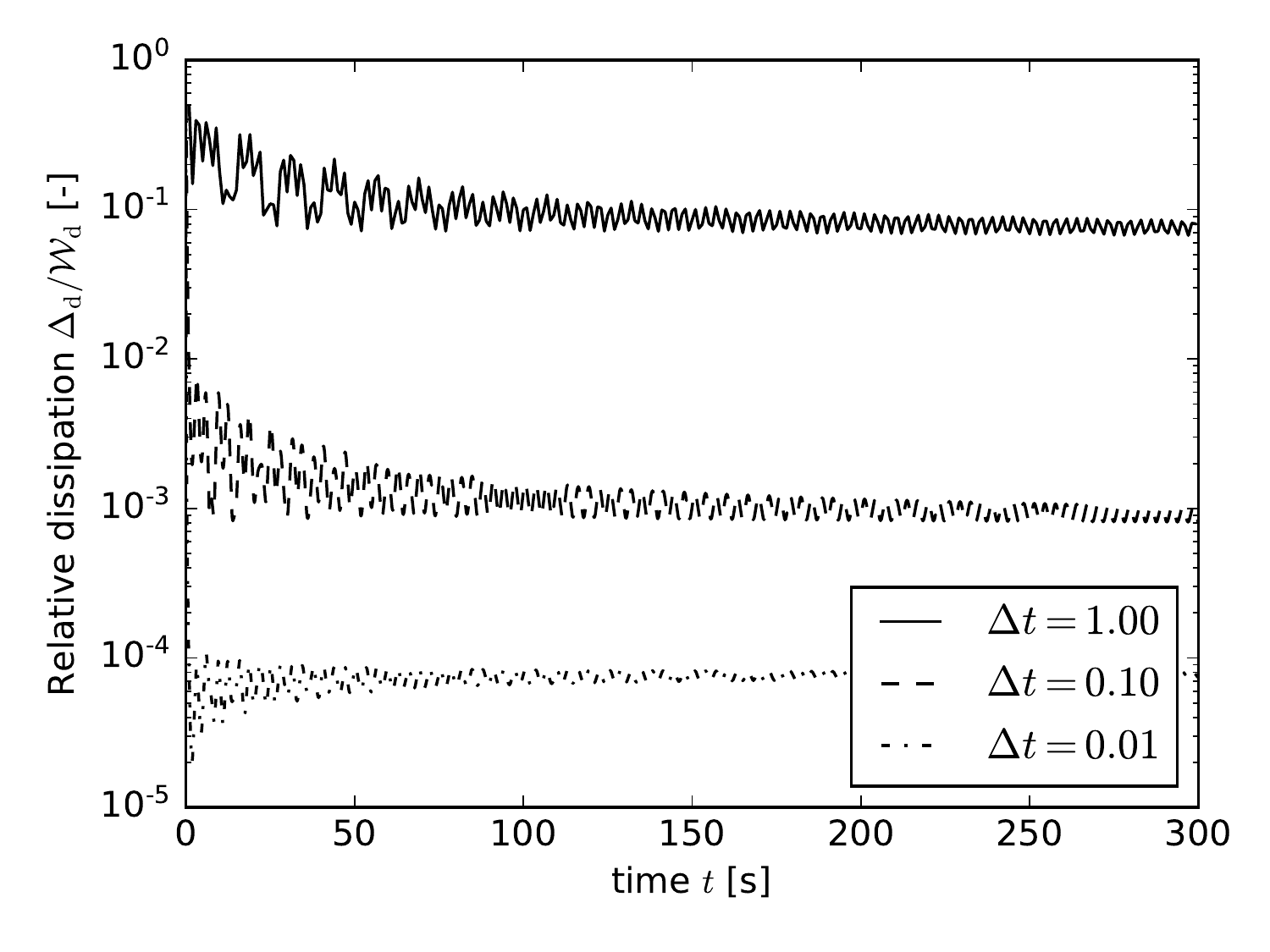} 
    \end{tabular}
    \caption{Normalized energies corresponding to SDOF response to step~(top) and harmonic loads~(bottom). Left: the evolution of internal energy $\mathcal{E}_\mathrm{int}$ and dissipation $\mathcal{D}\discr$, normalized by the work done by external forces $\mathcal{W}\discr$ for time step $\Delta t = 0.5$~s. Right: the evolution of numerical dissipation $\Delta\discr$, normalized by the work done by external forces~$\mathcal{W}\discr$.}
    \label{fig:results2}
\end{figure*}

\paragraph{Numerical dissipation.} As follows from the energy equality~\eqref{eq:energy_equality}, the additional dissipation induced by the integrator can be estimated as
\begin{align}\hspace{-.1em}
\Delta\discr(t_i)
=
\left|
\mathcal{E}_\mathrm{int}(0)
+
\mathcal{W}\discr(t_i)
- 
\mathcal{E}_\mathrm{int}(t_i)
-
\mathcal{D}\discr(t_i)
\right|,
\end{align}
with the individual terms provided by Eq.~\eqref{eq:discretized_energies}. The evolution of these quantities for the step and harmonic loading appears in \figref{fig:results2}, considering the time interval $\langle 0, 300 \rangle$~s.  

For both loads, we observe that the work done by external forces eventually distributes between the internal and dissipated energies; the ratio $\mathcal{D}\discr / \mathcal{W}\discr$ stabilizes at $0.4$ for the step load and for harmonic loading the ratio reaches about $0.9$. The artificial dissipation is only significant for the coarsest step of $\Delta t = 1.0$~s; for $\Delta t = 0.1$~s it reaches the value of about~$1$~\textperthousand~and further deteriorates with a decreasing time step. This confirms excellent energy conservation properties of the scheme, especially when taking into account that the error introduced by the trapezoidal rule in~\eqref{eq:discretized_energies} is of order $\mathcal{O}(\Delta t^2)$. 

\subsection{Generalization}\label{sec:example_continuum}
Additional constitutive assumptions must be adopted to extend the SDOF models into a continuum formulation. Here, we assume that the Maxwell model applies when modeling the response under shear. The spring stiffnesses~$k_\bullet$ thus become shear moduli $G_\bullet$, and that the Poisson ratio $\nu$ is a time-independent material constant. This assumption considerably simplifies the multi-dimensional constitutive law, e.g.,~\cite[Section~2.4]{bavzant2018creep}, and provides the same results as the conventional volumetric-deviatoric split for our target applications~\cite{zemanova_comparison_2017}. 

Under these assumptions, the weak form of the equations of motion attains the form, e.g.~\cite[Part~III]{clough2003dynamics}:
\begin{align}\label{eq:3D_motion}
    &
    \delta \mathcal{F}_\mathrm{ext}( t, \delta \uV)
    =
    \int_\Omega
    \delta\uV \cdot 
    \rho
    \ddot{\uV}(t)  
    \de\Omega
    \nonumber \\
    + & 
    \int_\Omega
    \delta\eps 
    : 
    \bigl( 
    G_\infty
    \bm{D}_\nu
    :
    \eps(t)
    +
    \sum_{p=1}^P
    \bm{\sigma}_p( t )
    \bigr)
    \de\Omega,
\end{align}
where $\delta \mathcal{F}_\mathrm{ext}$ stands for the virtual work done by external loads on a virtual displacement $\delta \uV$, $\ddot{\uV}$ denotes the acceleration, and the small strain tensor $\eps$ is obtained as the symmetric part of displacement gradient, $\eps = \nablas \uV$; virtual strain $\delta\eps$ is defined in the same way. The material is characterized by its density $\rho$, long-term shear modulus of the Maxwell chain $G_\infty$, and the dimensionless tensor $\bm{D}_\nu$ corresponding to the stiffness tensor of an isotropic material of unit shear modulus and the Poisson ratio of $\nu$. The stresses $\sig_p$ carried by individual cells follow as the solution of initial value problems
\begin{align}\label{eq:3D_internal}
    \frac{\dot{\sig}_p( t )}{G_p} 
    +
    \frac{\sig_p( t )}{\eta_p}
    = 
    \bm{D}_\nu \dot{\eps}(t),
    &&
    p = 1, 2, \ldots, P.
\end{align}
The evolution of the state variables $\uV$ and $\sig_p$ is specified with the initial conditions on the displacements, velocities, and cell stresses:
\begin{align}\label{eq:3D_initial}
\uV( 0 ) = \uV_0, 
&&
\dot{\uV}( 0 ) = \bm{v}_0,
&&
\sig_p( 0 ) = \sig_{p,0}.
\end{align}

The comparison of the initial value problems specified with Eqs.~\eqref{eq:equilibrium}, \eqref{eq:rdvisco}, and \eqref{eq:initial_conditions} and Eqs.~\eqref{eq:3D_motion}, \eqref{eq:3D_internal}, and~\eqref{eq:3D_initial} reveals that the derivation of the Newmark-type scheme follows exactly the steps as in \secref{sec:discretization}. As a result, the following variational problem needs to be solved at time $t_{i+1}$:
\begin{align}
&
\int_\Omega
\delta\uV \cdot 
\rho
\ddot{\uV}_{i+1}  
\de\Omega
\nonumber \\
+ &  
\int_\Omega
\delta\eps 
:
\Bigl(
    \quar G_\infty \Delta t^2 + \sum_{p=1}^P B_p
\Bigr)
\bm{D}_\nu
:
\ddot{\eps}_{i+1} 
\de\Omega
\nonumber \\
= & \;
\delta \mathcal{F}_\mathrm{ext}( t_{i+1}, \delta \uV) 
-
\int_\Omega
\delta\eps 
:
\Bigl(
    \sum_{p=1}^P A_p \sig_{p,i} 
\Bigr)
\de\Omega
\nonumber \\
- & 
\int_\Omega
\delta\eps 
:
G_\infty 
\bm{D}_\nu
:
\eps_i
\de\Omega 
\nonumber \\
- & 
\int_\Omega
\delta\eps 
:
\Bigl(
    G_\infty \Delta t 
    + 
    \sum_{p=1}^P G_p \widehat{\theta}_p
\Bigr)
\bm{D}_\nu
:
\dot{\eps}_{i} 
\de\Omega
\nonumber \\
- &
\int_\Omega
\delta\eps 
:
\Bigl(
    \quar G_\infty \Delta t^2 + \sum_{p=1}^P B_p
\Bigr)
\bm{D}_\nu
:
\ddot{\eps}_{i}
\de\Omega,
\label{eq:weak_form_Newmark}
\end{align}
with the parameters $\widehat{\theta}_p$, $A_p$, and $B_p$ provided by Eqs.~\eqref{eq:eff_relaxation} and~\eqref{eq:auxiliary_factors}; recall that $\delta \mathcal{F}_\mathrm{ext}$ denotes the virtual work done by external forces. Once the the accelerations $\ddot\uV_{i+1}$ are obtained from the weak form~\eqref{eq:weak_form_Newmark}, the displacements $\uV_{i+1}$, velocities $\dot{\uV}_{i+1}$, and the cell stresses $\sig_{p,i+1}$ are updated according to Eqs.~\eqref{eq:Newmark_kinematics} and~\eqref{eq:fvfinal2}, respectively. 

\begin{figure*}[h!]
\def\svgwidth{0.95\textwidth}
\begingroup%
  \makeatletter%
  \providecommand\color[2][]{%
    \errmessage{(Inkscape) Color is used for the text in Inkscape, but the package 'color.sty' is not loaded}%
    \renewcommand\color[2][]{}%
  }%
  \providecommand\transparent[1]{%
    \errmessage{(Inkscape) Transparency is used (non-zero) for the text in Inkscape, but the package 'transparent.sty' is not loaded}%
    \renewcommand\transparent[1]{}%
  }%
  \providecommand\rotatebox[2]{#2}%
  \newcommand*\fsize{\dimexpr\f@size pt\relax}%
  \newcommand*\lineheight[1]{\fontsize{\fsize}{#1\fsize}\selectfont}%
  \ifx\svgwidth\undefined%
    \setlength{\unitlength}{1087.8383381bp}%
    \ifx\svgscale\undefined%
      \relax%
    \else%
      \setlength{\unitlength}{\unitlength * \real{\svgscale}}%
    \fi%
  \else%
    \setlength{\unitlength}{\svgwidth}%
  \fi%
  \global\let\svgwidth\undefined%
  \global\let\svgscale\undefined%
  \makeatother%
  \begin{picture}(1,0.50642345)%
    \lineheight{1}%
    \setlength\tabcolsep{0pt}%
    \put(0,0){\includegraphics[width=\unitlength,page=1]{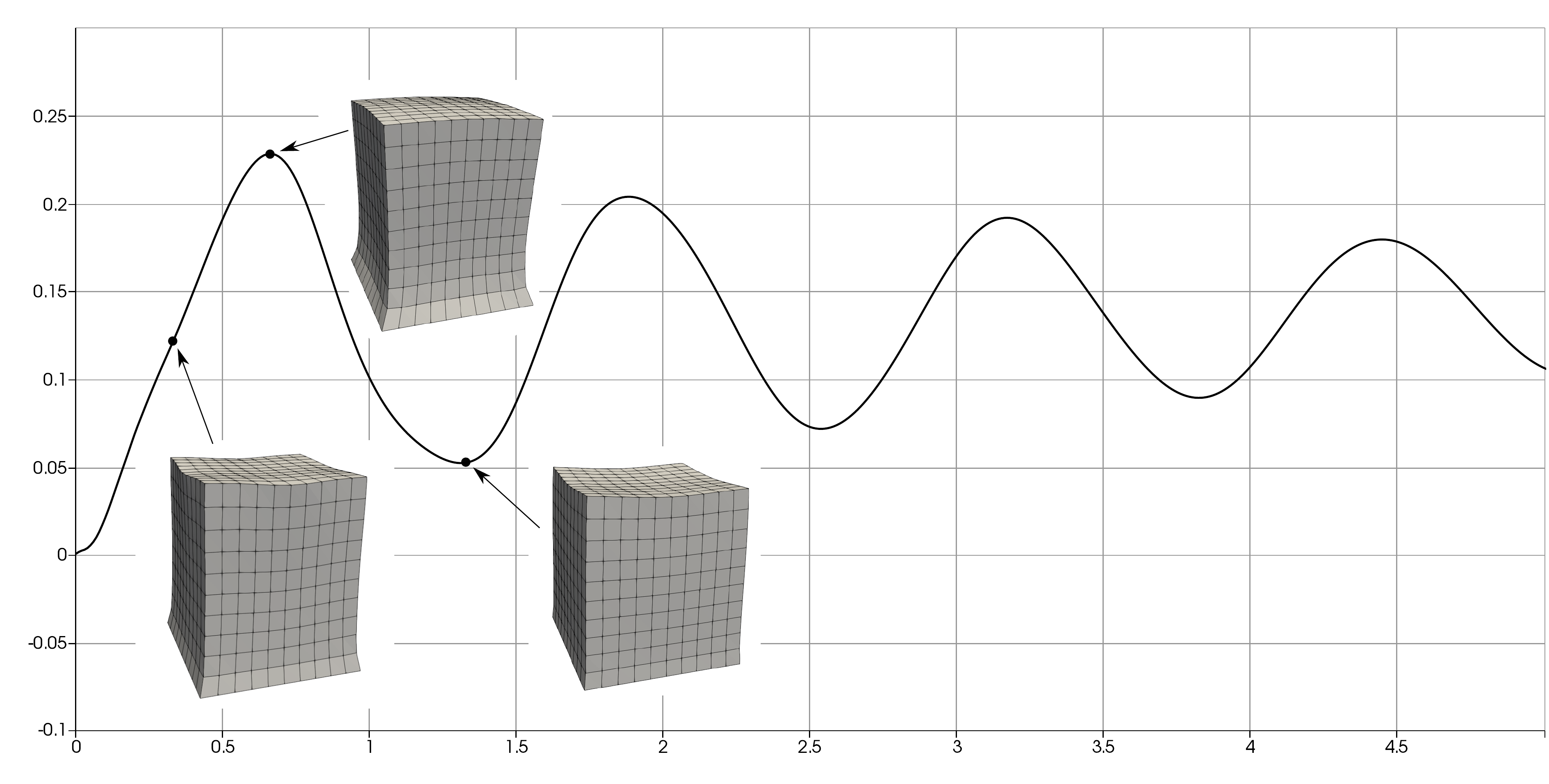}}%
    \put(0.48810366,0.0016466){\color[rgb]{0,0,0}\makebox(0,0)[lt]{\lineheight{1.25}\smash{\begin{tabular}[t]{l}time $t$ [s]\end{tabular}}}}%
    \put(0.00851197,0.19370516){\color[rgb]{0,0,0}\rotatebox{90}{\makebox(0,0)[lt]{\lineheight{1.25}\smash{\begin{tabular}[t]{l}displacement $r(t) [m]$\end{tabular}}}}}%
    \put(0,0){\includegraphics[width=\unitlength,page=2]{figure04.pdf}}%
  \end{picture}%
\endgroup%

\caption{Snapshots of deformations of a viscoelastic cube subjected to ramp load.}
\label{fig:3dexample}
\end{figure*}

The outlined formulation~\eqref{eq:weak_form_Newmark} was further discretized with the finite element method and implemented in FEniCS project~\cite{logg_automated_2012,AlnaesBlechta2015a} version 2018.1.  As an indicative example, we consider a unit cube, see \figref{fig:3dexample}, fixed on the bottom surface and subjected to a ramp load~\eqref{eq:ramp_load} with the tensile traction of intensity 1.0~Nm$^{-2}$ perpendicular to the top surface. The material response is characterized by the Maxwell chain parameters from \tabref{tab:pronySeries} and the value of the Poisson ratio $\nu = 0.49$. In the numerical resolution, we discretize the sample into identical 1,000 hexahedron elements and consider the time step of 0.01~s. 

The snapshots of the vibrations reveal a similar behavior to the SDOF example, recall \figref{fig:sdof_results}, namely the attenuation of the propagating waves by viscous damping. An interested reader is invited to the dataset~\cite{schdmit_jaroslav_2019_3265058} for full details on the simulation. 

\section{Conclusions}\label{sec:conc}

In this contribution, we have developed a Newmark integration scheme for viscoelastic solids characterized by the generalized Maxwell model. Besides the direct derivation, we have shown the scheme can be derived from the Hamilton variational principle combined with a suitable structure-preserving time discretization. This variational structure is then reflected in the long-term stability and low energy dissipation of the resulting scheme, which has been confirmed with selected numerical examples. 

As the next step, we will combine the continuum framework outlined in \secref{sec:example_continuum} with Newmark-type solvers for variational fracture models, e.g.~\cite{li_numerical_2016,Roubicek2019}, to extend the currently available approaches to simulating the response of laminated glass structures under impact.
    
\begin{acknowledgements}
This publication was supported by the Czech Science Foundation, the grant No.~19-15326S.
\end{acknowledgements}

\end{document}